\documentclass[prl,preprintnumbers,twocolumn,showpacs,nobalancelastpage,superscriptaddress]{revtex4}

\usepackage{amsmath,amssymb,amsfonts}
\usepackage{bm,graphicx,color}

\newcommand{\CC}{\mathcal{C}}

\newcommand{\TR}{\text{Tr}}
\newcommand{\BK}{{\bm{k}}}
\newcommand{\mydagger}{{+}}
\newcommand{\phdagger}{{\phantom{+}\!}}
\newcommand{\ETAL}{{\em et al.}}

\newcommand{\bra}[1]{\langle{#1}|}
\newcommand{\ket}[1]{|{#1}\rangle}

\newcommand{\expval}[1]{\langle{#1}\rangle}

\newcommand{\ret}{\text{ret}}

\newcommand{\Kplus}{K_+}
\newcommand{\Kminus}{K_-}

\newcommand{\uc}{{3.2}}

\begin{document}

  \title{Dynamical phase transition in correlated fermionic lattice systems}

  \author{Martin Eckstein}
  
  \author{Marcus Kollar}
  
  \affiliation{Theoretical Physics III, Center for Electronic
    Correlations and Magnetism, Institute for Physics, University of
    Augsburg, 86135 Augsburg, Germany}

  \author{Philipp Werner}
  
  \affiliation{Theoretical Physics, ETH Zurich, 8093 Zurich, Switzerland}
  
  \date{April 6, 2009}

  \begin{abstract}
    We use non-equilibrium dynamical mean-field theory to demonstrate
    the existence of a critical interaction in the real-time dynamics
    of the Hubbard model after an interaction quench. The \makebox{critical}
    point is characterized by fast thermalization and separates
    weak-coupling and strong-coupling regimes in which the relaxation
    is delayed due to prethermalization on intermediate timescales.
    This dynamical phase transition should be observable in
    experiments on trapped fermionic atoms.
  \end{abstract}


  \maketitle
  
  The properties of correlated many-particle systems can change
  dramatically and abruptly as external parameters are varied. An
  important example is the Mott transition from an itinerant state to
  a correlation-induced insulator, which occurs in such diverse
  systems as transition-metal compounds~\cite{Imada1998a} and
  ultracold quantum
  gases~\cite{Greiner2002,Joerdens2008a,Schneider2008a}.  An entirely
  new perspective on those systems is provided by their nonequlibrium
  dynamics after an external perturbation, which is experimentally
  accessible not only in the case of well-controlled ultracold quantum
  gases, but also for electrons in solids by means of femtosecond
  spectroscopy~\cite{Ogasawara00,Iwai03,Perfetti2006,Okamoto07,Kuebler07}.
  On short timescales the perturbed systems are essentially decoupled
  from the environment and follow the unitary time evolution according
  to the Schr\"{o}dinger equation, which immediately raises a number
  of questions: How does an isolated many-body system approach a new
  equilibrium after being quenched, i.e., after a sudden change in one
  of its parameters? Does it eventually thermalize, or is detailed
  memory on the initial state retained for all times?
  
  Recently these questions have been addressed in a number of
  experimental~\cite{Kinoshita06,Hofferberth2007a} and theoretical
  investigations~\cite{Rigol07+06,Cazalilla06,Kollath07, Manmana07,
    Rigol2008,Eckstein2008,Moeckel2008,Kollar2008,Rossini2009a,
    Barmettler2008a}. After a quench to a large interaction parameter
  $U$ characteristic collapse-and-revival oscillations with period
  $2\pi\hbar/U$ appear, which are due to the integer eigenvalues of
  the interaction
  operator~\cite{Greiner2002,Misochko2004a,Rigol07+06,Kollath07,%
    Manmana07,Eckstein2008,Kollar2008,Rossini2009a,Barmettler2008a}.
  These oscillations eventually fade out, but in some cases the system
  is trapped in a nonthermal stationary state up to the largest
  accessible times~\cite{Kollath07,Manmana07}.  At first glance this
  may seem surprising because thermalization is only known to be
  inhibited for integrable
  systems~\cite{Rigol07+06,Cazalilla06,Kollar2008}, whereas
  nonintegrable systems such as those studied in
  Refs.~\onlinecite{Kollath07,Manmana07} are expected to
  thermalize~\cite{Rigol2008}.  Nonthermal quasistationary states are
  found, on the other hand, on an intermediate timescale
  ($\propto1/U^2$) after quenches to small interaction parameters
  ~\cite{Moeckel2008}; the prethermalization to these states is
  followed by an approach to thermal equilibrium on a much longer
  timescale ($\propto1/U^4$).  Another interesting observation is that
  the relaxation dynamics can depend sensitively on the parameters of
  the final Hamiltonian~\cite{Barmettler2008a}.  Whether and how this
  phenomenon, which may be called a dynamical phase transition,
  relates to the existence of an equilibrium thermodynamic phase
  transition, remains to be clarified.

  Here we consider the relaxation of correlated lattice fermions
  described by a time-dependent Hubbard Hamiltonian at half-filling,
  \begin{align}
    \label{hubbard}
    H(t)
    &=
    \sum_{ij\sigma}
    V_{ij}
    c_{i\sigma}^{\mydagger}
    c_{j\sigma}^{\phdagger}
    +
    U(t)
    \sum_{i}
    \big(n_{i\uparrow}\!-\!\tfrac12\big)
    \big(n_{i\downarrow}\!-\!\tfrac12\big)
    \,,
  \end{align}
  using nonequilibrium dynamical mean-field theory (DMFT).  We
  restrict ourselves to the paramagnetic phase and choose hoppings
  $V_{ij}$ corresponding to a semi-elliptic density of states
  $\rho(\epsilon)$ $=$ $\sqrt{4V^2-\epsilon^2}/(2\pi V)$.  The system
  is initially in the ground state of the noninteracting Hamiltonian,
  i.e., $U(t$$<$$0)$ $=$ $0$.  At $t$ $=$ $0$ the Coulomb repulsion is
  switched to a finite value, $U(t$$\ge$$0)$ $=$ $U$.  Energy is
  measured in units of the quarter-bandwidth $V$ and time in units of
  $1/V$, i.e., we set $\hbar$ $=$ $1$ and in the figures also $V$ $=$
  $1$. Our results confirm the prethermalization for quenches to $U$
  $\ll$ $V$ and indicate a second prethermalization regime for $U$
  $\gg$ $V$, for which we provide a general perturbative argument.  At
  $U_c^{\text{dyn}}$ $=$ $\uc V$ we observe a sharp crossover between
  the two regimes, suggestive of a dynamical phase transition in the
  above sense.

  {\em Nonequilibrium DMFT.---} %
  In equilibrium DMFT, which becomes exact in the limit of infinite
  dimensions~\cite{Metzner1989}, the self-energy is local and can be
  calculated from a single-site impurity model subject to a
  self-consistency condition~\cite{Georges1996}.  Nonequilibrium DMFT
  is a reformulation for Green functions on the Keldysh
  contour~\cite{Schmidt2002,Freericks2006}, which maps the lattice
  problem (\ref{hubbard}) onto a single-site problem described by the
  action
  \begin{align}
    \label{action}
    \mathcal{S}
    =
    \sum_{\sigma=\uparrow,\downarrow}
    \int_\CC \!\!dt\,dt'\,c_\sigma^\mydagger(t)
    \Lambda_\sigma(t,t')c_\sigma(t')
    +\int_\CC\!dt\, h_{\text{loc}}(t),
  \end{align}
  where $h_{\text{loc}}(t)$ $=$
  $U(t)$$[n_\uparrow(t)$$-$$\frac{1}{2}]$$[n_\downarrow(t)$$-$$\frac{1}{2}]$
  is the local Hamiltonian at half-filling. For a system that is
  prepared in equilibrium with temperature $T$ at times $t$ $<$ $0$
  the time-contour $\CC$ is chosen to run from $t$ $=$ $0$ along the
  real axis to $t_{\text{max}}$, back to $0$, and finally to $-i/T$
  along the imaginary axis. The action $\mathcal{S}$ determines the
  contour-ordered Green function $G_{\sigma}(t,t')$ $=$ $\TR[ T_\CC
  e^{-i\mathcal{S}} c_{\sigma}(t) c_{\sigma}^\mydagger(t')]/Z$ and the
  self-energy $\Sigma(t,t')$.  For the semi-elliptic density of states
  the selfconsistency condition reduces to $\Lambda_\sigma(t,t')$ $=$
  $V^2 G_\sigma(t,t')$~\cite{Eckstein2008}.  The single-site problem
  can be solved, for example, by means of real-time Monte Carlo
  techniques~\cite{Muehlbacher2008, Werner2009}. We use the
  weak-coupling continuous-time Monte Carlo (CTQMC)
  algorithm~\cite{Werner2009}, which stochastically samples a
  diagrammatic expansion in powers of the interaction part
  $h_{\text{loc}}$ and measures observables such as the local Green
  function $G_\sigma(t,t')$.  The weak-coupling method is highly
  suitable for initial noninteracting states, because the imaginary
  branch of the contour does not enter the CTQMC calculation. This
  allows us to study initial states at zero temperature by
  transforming imaginary times to real frequencies.  Furthermore, the
  parameters of the algorithm can be chosen such that only even orders
  contribute to $G_\sigma(t,t')$ at half-filling, which reduces the
  sign problem.  Computational details are deferred to a separate
  publication.

  Many observables of the lattice model can be calculated from the
  local Green function $G_\sigma(t,t')$ and the self-energy
  $\Sigma_\sigma(t,t')$.  The two-particle correlation function
  $\Gamma_\sigma(t,t')$ $=$ $\expval{T_\CC
    c_{i\sigma}^\mydagger(t)[n_{i\bar
      \sigma}(t)-\frac{1}{2}]c_{i\sigma}(t')}$ is obtained from the
  contour convolution $\Gamma_\sigma$ $=$ $G_\sigma$ $\ast$
  $\Sigma_\sigma$ and yields the double occupation $d(t)$ $=$
  $\expval{n_{i\uparrow}(t) n_{i\downarrow}(t)}$ at $t$ $=$ $t'$.
  (Local quantities do not depend on the site index $i$ for the
  homogeneous phase.) Solving the the lattice Dyson equation
  $(i\partial_t + \mu -\epsilon -\Sigma_\sigma)$ $\ast$
  $G_{\epsilon\sigma}$ $=$ $1$ yields the momentum-resolved equal-time
  Green function $G_{\epsilon\sigma}(t,t)$ and thus the momentum
  distribution $n(\epsilon_{\BK},t)$ $=$
  $\expval{c_{\BK\sigma}^\mydagger(t)c_{\BK\sigma}^\phdagger(t)}$, as
  well as the kinetic energy per lattice site, $E_{kin}(t)/L$ $=$
  $2\int\!d\epsilon\,\rho(\epsilon)\,n(\epsilon,t)\,\epsilon$.  The
  total energy $E$ $=$ $E_{kin}(t)+UL\,[d(t)-1/4]$ must be conserved
  after the quench, and we use this as a check for the numerical
  solution.

  \begin{figure}
    \centerline{\includegraphics[width=100mm]{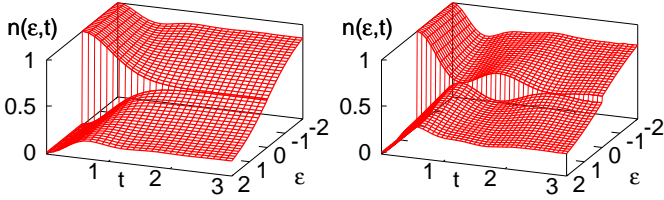}}
     \caption{Momentum distribution $n(\epsilon_{\BK},t)$ for
      quenches from $U$ $=$ $0$ to $U$ $=$ $3$ (left panel) and $U$
      $=$ $5$ (right panel).}
    \label{fig:nku5}
  \end{figure}

  {\em Results.---} %
  As depicted in Fig.~\ref{fig:nku5}, the momentum distribution
  $n(\epsilon,t)$ evolves from a step function in the initial state to
  a continuous function of $\epsilon$ at large times.  Remarkably, its
  discontinuity at $\epsilon$ $=$ $0$, which marks the Fermi surface
  in the initial state, remains sharp while its height decays smoothly
  to zero.  For a noninteracting initial state at half-filling, the
  discontinuity $\Delta n(t)$ $=$ $n(0^-,t)-n(0^+,t)$ can be expressed
  as
  \begin{align}
    \label{jump}
    \Delta n(t) = |G_{\epsilon=0,\sigma}^\ret(t,0)|^2
    \,,
  \end{align}
  where $G_{\epsilon_{\BK}\sigma}^\ret(t,0)$ $=$
  $-i\Theta(t)\expval{\{c^\phdagger_{\BK\sigma}(0),c^\mydagger_{\BK\sigma}(t)\}}$
  is the retarded component of the momentum-resolved Green function.
  This shows that the collapse of the discontinuity $\Delta n$ is
  closely related to the decay of electron and hole excitations which
  are created at time $t$ $=$ $0$ at the Fermi surface.
  
  We now use $\Delta n(t)$ and $d(t)$ to characterize the relaxation
  after the quench. As shown in detail below, these functions behave
  qualitatively different in the weak-coupling and strong-coupling
  regimes, separated by a very sharp crossover at $U_c^{\text{dyn}}$
  $=$ $\uc V$.  We test for thermalization by comparing to expectation
  values in a grand-canonical ensemble with the same total energy.

  \begin{figure}
    \includegraphics[width=86mm]{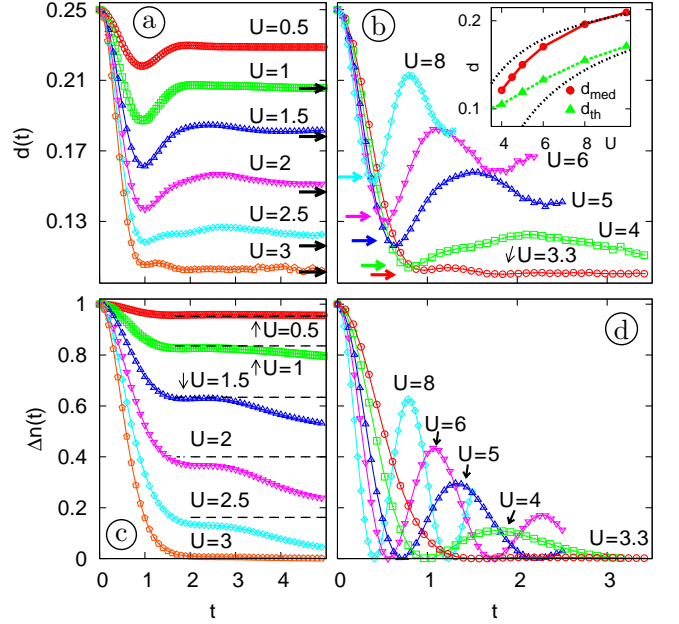}
    \caption{Fermi surface discontinuity $\Delta n$ and double
      occupation $d(t)$ after quenches to $U\le3$ (left panels) and
      $U\ge 3.5$ (right panels).  Horizontal dotted lines in the upper
      left panel are at the quasistationary value $\Delta
      n_{\text{stat}}$ $=$ $2Z-1$ predicted in
      Ref.~\onlinecite{Moeckel2008}, with the $T=0$ quasiparticle
      weight $Z$ taken from equilibrium DMFT data~\cite{Bulla1999}.
      Horizontal arrows indicate corresponding thermal values
      $d_{\text{th}}$ of the double occupation, obtained from
      equilibrium DMFT using QMC. Inset: thermal value
      $d_{\text{th}}$ and $d_{med}$, the average of the first maximum
      and the second minimum of $d(t)$, which provides an estimate of
      the stationary value $d_{\text{stat}}$; black dotted lines are
      the respective results from the strong-coupling expansion (see
      text).}
    \label{fig:double}
  \end{figure}

  {\em Weak-coupling regime, $U$ $<$ $U_c^{\text{dyn}}$.---} %
  For quenches to $U$ $\le$ $3V$ (Fig.~\ref{fig:double}a,c) we find
  that the double occupation $d(t)$ relaxes from its initial
  uncorrelated value $d(0)$ $=$
  $\expval{n_\uparrow}_0\expval{n_\downarrow}_0$ $=$
  $1/4$~\cite{zerosubscript} almost to its thermal value
  $d_{\text{th}}$, while the Fermi surface discontinuity $\Delta n(t)$
  remains finite for times $t$ $\leq$ $5/V$.  This confirms the
  predictions by Moeckel and Kehrein~\cite{Moeckel2008} of a
  quasistationary state which is formed on timescales on the order of
  $V/U^2$ and has $d_{\text{stat}}$ $=$ $d_{\text{th}}$ $+$
  $\mathcal{O}(U^3/V^3)$ and finite $\Delta n_{\text{stat}}$ $=$
  $1-2Z$. Here $Z$ is the quasiparticle weight in equilibrium at zero
  temperature and interaction $U$.  Their prediction was based on a
  perturbative flow equation analysis for $U$ $\ll$ $V$ and it was
  argued that full thermalization occurs only on much longer
  timescales on the order of $V^3/U^4$.  At $t=2/V$ our numerical data
  agree very well with the predicted value of $\Delta n_{\text{stat}}$
  for $U$ $\leq$ $1V$.  Note that even for quenches to larger $U$, a
  prethermalization plateau remains visible in Fig.~\ref{fig:double}c
  at roughly this value, although the timescales $V/U^2$ and $V^3/U^4$
  are no longer well separated.

  {\em Strong-coupling regime, $U$ $>$ $U_c^{\text{dyn}}$.---} For
  quenches to large $U$ we observe collapse-and-revival oscillations
  with approximate frequency $2\pi/U$ both in $d(t)$ and $\Delta n(t)$
  (Fig.~\ref{fig:double}b,d).  This phenomenon is well understood in
  the atomic limit ($V$ $=$ $0$), where the propagator $e^{-iHt}$ is
  exactly $2\pi/U$-periodic~\cite{Greiner2002}.  As expected, these
  oscillations are damped for nonzero $V$: at least for small times,
  they fall off on timescales on the order of $1/V$.  Interestingly,
  the first few oscillations of $d(t)$ are not centered around the
  thermal value $d_{\text{th}}$ (solid arrows in
  Fig.~\ref{fig:double}b), which is instead located close to the first
  minimum of $d(t)$ This suggests a prethermalization regime also for
  $U$ $\gg$ $V$, as discussed in Ref.~\onlinecite{Kollath07}, where
  oscillations in $d(t)$ are damped to a nonthermal quasistationary
  value on the timescale $1/V$, while full thermalization can only
  happen on the longer timescale $U/V^2$.

  We now show that this prethermalization regime is a general feature
  of fermionic Hubbard-type models at strong coupling and calculate
  the double occupation in the quasistationary state.  We use the
  standard unitary transformation $\bar A$ $=$
  $e^{-S}A\,e^S$~\cite{Harris1967} for which the double occupation
  $\bar D$ $=$ $\sum_i\bar n_{i\uparrow}\bar n_{i\downarrow}$ of the
  dressed fermions $\bar c_{i\sigma}$ is conserved, $[H,\bar D]$ $=$
  $0$.  After decomposing the hopping term~\cite{footnotespin}, $K$
  $=$ $\sum_{ij\sigma} (V_{ij\sigma}/V) c_{i\sigma}^\mydagger
  c_{j\sigma}$, into parts $K_p$ that change the double occupation by
  $p$, i.e., $\Kplus$ $=$ $\sum_{ij\sigma} (V_{ij\sigma}/V)
  c_{i\sigma}^\mydagger c_{j\sigma}
  (1-n_{j\bar\sigma})n_{i\bar\sigma}$ $=$ $(\Kminus)^\mydagger$ and
  $K_0$ $=$ $K$ $-$ $\Kplus$ $-$ $\Kminus$, the leading order
  transformation is $S$ $=$ $(V/U)\bar \Kplus$ $+$ $(V/U)^2[\bar
  \Kplus,\bar K_0]$ $-$ $\text{h.c.}$ $+$ $\mathcal{O}(V^3/U^3)$.  For
  the double occupation, $d(t)$ $=$ $\expval{e^{iHt}De^{-iHt}}_0/L$,
  we obtain
  \begin{align}
    \label{d1}
    d(t) &= d_{\text{stat}}- \frac{2V}{U} \text{Re}\big[ e^{itU}
    R(tV)\big] + \mathcal{O}\Big(
    \frac{V^2}{U^2},\frac{tV^3}{U^2}\Big)\,,
  \end{align}
  where $R(tV)$ $=$ $\expval{e^{i t V K_0} \Kplus e^{-it V K_0}}_0/L$
  and $d_{\text{stat}}$ $=$ $d(0) +
  (2V/U)\text{Re}\expval{\Kplus/L}_0$. The error
  $\mathcal{O}(tV^3/U^2)$, which is due to omitted terms in the
  exponentials $e^{\pm iHt}$, is irrelevant in comparison to the
  leading terms if $t$ $\ll$ $U/V^2$.  It remains to show that (i)~the
  envelope function $R(tV)$ of the oscillating term decays to zero for
  $t$ $\gg$ $1/V$, and (ii)~the quasistationary value
  $d_{\text{stat}}$ differs from the thermal value $d_{\text{th}}$.
  (i)~Inserting an eigenbasis $K_0\ket{m}$ $=$ $k_m\ket{m}$ yields
  $R(tV)$ $=$ $\sum_{m,n} \langle \ket{n}\bra{m}\rangle_0
  e^{itV(k_m-k_n)}$ $\bra{n} \Kplus \ket{m}$.  In this expression all
  oscillating terms dephase in the long-time
  average~\cite{Rigol2008,Kollar2008}, so that only energy-diagonal
  terms contribute to the sum.  But from $[K_0,D]$ $=$ $0$ it follows
  that $D$ is a good quantum number of $\ket{n}$ so that $\bra{n}
  \Kplus\ket{n}$ $=$ $0$, and thus $R(tV)$ vanishes in the long-time
  limit (if it exists and if accidental degeneracies between sectors
  of different $D$ are irrelevant). From Eq.~(\ref{d1}) we therefore
  conclude that $d(t)$ equals $d_{\text{stat}}$ for times $1/V$ $\ll$
  $t$ $\ll$ $U/V^2$, up to corrections of order
  $\mathcal{O}(V^2/U^2)$.  (ii)~For the quasistationary value
  $d_{\text{stat}}$ we obtain
  \begin{subequations}%
    \label{dstat}%
    \begin{align}%
      d_{\text{stat}} &= d(0)-\Delta d
      \,,\\
      \label{deltad}%
      \Delta d &= - \sum_{ij\sigma}
      \frac{V_{ij\sigma}}{UL}\expval{c_{i\sigma}^\mydagger
        c_{j\sigma}(n_{i\bar\sigma}-n_{j\bar\sigma})^2}_0 \,,%
    \end{align}%
  \end{subequations}%
  which applies to arbitrary initial states. For noninteracting
  initial states the expectation value in this expression factorizes;
  in DMFT Eq.~(\ref{deltad}) then evaluates to $\Delta d$ $=$
  $n(1-n/2)(V/U)\expval{K/L}_0$, i.e., it is proportional to the
  kinetic energy in the initial state. For the thermal value
  $d_{\text{th}}$ we expand the free energy in $V/T_*$, because the
  effective temperature $T_*$ is much larger than $V$ after a quench
  to $U$ $\gg$ $V$.  At half-filling we obtain $d_{\text{th}}$ $=$
  $d(0)$ + $(V/U)\expval{K/L}_0$; for noninteracting initial states in
  DMFT we thus find that $\Delta d$ $=$ $d(0)-d_{\text{stat}}$ $=$
  $[d(0)-d_{\text{th}}]/2$, i.e., at times $1/V$ $\ll$ $t$ $\ll$
  $U/V^2$ the double occupation has relaxed only halfway towards
  $d_{\text{th}}$.

  The strong-coupling predictions for the prethermalization plateau
  agree with our numerical results, for which the center of the first
  oscillation in $d(t)$ approaches $d_{\text{stat}}$ for large $U$
  (inset in Fig.~\ref{fig:double}b).  The scenario also applies to
  interaction quenches in the half-filled Falicov-Kimball model in
  DMFT~\cite{Eckstein2008} and the $1/r$ Hubbard
  chain~\cite{Kollar2008}, although thermalization is inhibited in
  these models: in both models the long-time limit of
  $d(t$$\to$$\infty)$ can be obtained exactly and indeed agrees with
  $d_{\text{stat}}$ for $U$ $\gg$ $V$. For quenches to large $U$ in
  the free $1/r$ chain (with bandwidth $2\pi V$) Eq.~(\ref{deltad})
  yields $\Delta d$ $=$ $(V/U)(1-2n/3)\pi$.  For the Falicov-Kimball
  model $\Delta d$ is half as big as for the Hubbard model because
  only one spin species contributes to the kinetic energy in the
  initial state.

  \begin{figure}
    \centerline{\includegraphics[width=88mm]{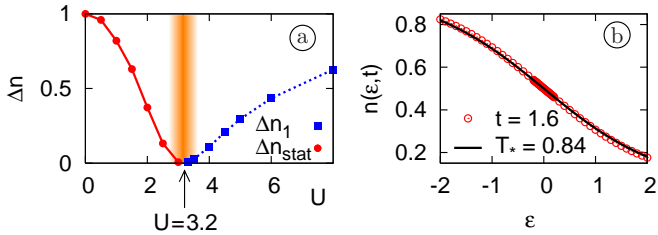}}
    \caption{Left panel: $U$-dependence of the prethermalization
      plateau, $\Delta n_\text{stat}$, and the first revival maximum,
      $\Delta n_1$, suggesting a dynamical phase transition at
      $U_c^{\text{dyn}}$ $=$ $\uc V$. Right panel: momentum
      distribution at time $t$ $=$ $1.6$ after a quench to $U$ $=$
      $3.3V$; the momentum distribution has essentially thermalized.
      The shaded area marks the critical region near $U_c^\text{dyn}$
      $=$ $\uc V$ with very fast thermalization.}
    \label{fig:crit}
  \end{figure}

  {\em Critical region, %
    $U$ $\approx$ $U_c^\text{dyn}$ $=$ $\uc V$.---} %
  The characteristic col\-lapse-and-revival oscillations of the
  strong-coupling regime disappear for quenches to $U$ between $3.3V$
  and $3V$, as is apparent from the Fermi surface discontinuity
  $\Delta n_1$ at its first revival maximum (Fig.~\ref{fig:crit}a).
  This change in the short-time dynamics reflects a change in the
  nature of single-particle excitations [Eq.~(\ref{jump})]. It occurs
  also in equilibrium even at very high temperatures, because
  $|G^\ret_{\epsilon\sigma}(t-t')|^2$ becomes oscillatory upon the
  transfer of spectral weight to the Hubbard subbands at $\pm U$.
  Additionally the prethermalization plateau at $\Delta n_\text{stat}$
  disappears between $3V$ and $3.3V$, so that the system can relax
  rapidly after quenches to $U$ values in this range: momentum
  distribution and double occupation reach their respective thermal
  values already before the first expected collapse-and-revival
  oscillation at time $2\pi/U$ (Fig.~\ref{fig:crit}b).  This suggests
  that the prethermalization regimes at weak and strong Hubbard
  interaction are separated by a special point, which we estimate
  from our data to be located at $U_c^\text{dyn}$ $=$ $3.2V$, where 
  relaxation processes on all energy scales become relevant.
  This sharp crossover is quite remarkable in view of the fact that
  the effective temperature $T_*$ after the quench is much higher than
  the critical endpoint of the Mott metal-insulator transition in
  equilibrium ($T_c \approx 0.055V$~\cite{Georges1996}, whereas $T_*$
  $=$ $0.84V$ for $U$ $=$ $3.3V$), so that in equilibrium metallic and
  insulating phases could hardly be distinguished. A similar critical
  behavior was found for quenches in Heisenberg
  chains~\cite{Barmettler2008a}. Therefore one might speculate that
  such {\em dynamical phase transitions} are a generic property of the
  nonequilibrium dynamics of correlated systems; however, this issue
  requires further study. In particular it would be interesting to see
  whether there is also an abrupt change in the long-time behavior at
  the same $U_c^{\text{dyn}}$.
  
  {\em Conclusion.---} %
  We determined the real-time dynamics of the Hubbard model after a
  quench from the noninteracting state using nonequilibrium DMFT, for
  which CTQMC~\cite{Muehlbacher2008,Werner2009} proves to be a very
  suitable impurity solver.  This method allows to investigate the
  transient dynamics of correlated fermions in a variety of contexts,
  e.g., to describe experiments with cold atomic
  gases~\cite{Joerdens2008a,Schneider2008a} or pump-probe spectroscopy
  on correlated electrons~\cite{Eckstein2008}.  For the Hubbard model
  we found that rapid thermalization occurs after quenches to $U$
  $\approx$ $U_c^{\text{dyn}}$; in fact this is one of the few
  cases~\cite{Kollath07,Rigol2008} where thermalization can be
  numerically observed for a non-integrable model.  On the other hand,
  for quenches to very small or very large interactions, the system
  becomes trapped in quasistationary states on intermediate
  timescales.  The phenomena discussed in this paper are manifest in
  the momentum distribution and double occupation; both
  quantities are directly accessible in experiments with cold atomic
  gases.

  We thank S.\ Kehrein, C.\ Kollath, M.\ Moeckel, M.\ Punk, M.\ Rigol,
  A.\ Silva, and D.\ Vollhardt for useful discussions.  
  M.E.\ acknowledges support by Studienstiftung des deutschen Volkes.  
  This work was supported in part by SFB 484 of the Deutsche
  Forschungsgemeinschaft (M.E., M.K.)  and grant PP002-118866 of the
  Swiss National Science Foundation (P.W.).  CTQMC calculations were
  run on the Brutus cluster at ETH Zurich, using the ALPS
  library~\cite{ALPS}.

  %%%%%%%%%%%%%%%%%%%%%%%%%%%%%%%%%%%%%%%%%%%%%%%%%%%%%%%%%%%%% 

\end{document}